# *Ab initio* intermolecular interactions mediate thermochemically real-fluid effects that affect system reactivity

Mingrui Wang, Ruoyue Tang, Xinrui Ren, Yanqing Cui, Song Cheng[*]

*Department of Mechanical Engineering, The Hong Kong Polytechnic University, Hung Hom, Kowloon, Hong Kong*

______________________________________________________________________

**Abstract**

The properties of supercritical fluids are dictated by intermolecular interactions that involve two or more molecules. Such intermolecular interactions were described via intermolecular potentials in historical supercritical combustion modeling studies, but have been treated empirically and with no consideration of radical interactions or multi-body interactions involving more than two molecules. This approach has been adopted long ago, assuming sufficient characterization of real-fluid effects during supercritical combustion. Here, with data from *ab initio* multi-body intermolecular potentials, non-empirical Virial Equation of State (EoS), and real-fluid thermochemical and kinetic simulations, we reveal that empirical intermolecular potentials can lead to significant errors in representing supercritical fluids under common combustion situations, which can be impressively described by *ab initio* intermolecular potentials. These interactions are also found to greatly influence autoignition delay times, a common measure of global reactivity, with significant contributions from radical interactions and multi-body interactions. It is therefore of necessity to incorporate *ab initio* intermolecular interactions in studying supercritical combustion and various dynamic systems involving supercritical fluids, which has now been enabled through the new framework developed in the present study.

______________________________________________________________________

*Corresponding author.
Email: songryan.cheng@polyu.edu.hk
Song Cheng
Phone: +852 2766 6668

The mounting concerns over the emissions and environmental impacts of the transportation sector have provoked urgent demand for high-pressure combustion and propulsion technologies due to its high power output, high thermal efficiency, high fuel economy and low emissions [1, 2]. This is particularly imminent for heavy-duty aviation and aerospace transportation systems where power density is highly sought-after and electrification via batteries is not viable. As a result, the past three decades have witnessed an exponential rise in the power density of various combustion and propulsion systems. Today, advanced on-road heavy-duty engines, gas turbines and rocket engines can easily operate at 100–300 atm, where the system working fluid becomes supercritical.

Under supercritical conditions, intermolecular interactions in the working fluid are greatly enhanced, while molecular collisional cross section and collision frequency alter considerably [3]. Due to these changes, the system molecular distribution deviates from a Boltzmann distribution, and the multi-body (>2) interactions that are not important at subcritical conditions become prominent. These non-ideal molecular behaviors eventually manifest themselves in fluid properties, resulting in the unique properties of supercritical fluids. For instance, supercritical fluids exhibit liquid-like density and gas-like diffusivity [4] with zero latent heat and large compressibility [5]. Due to these shifts in fluid properties, supercritical combustion exhibits dramatically different characteristics than those typically observed at subcritical conditions. Therefore, our ability to wield and exploit complex supercritical systems, such as supercritical propulsion engines and supercritical hydrogen storage systems, is premised on our quantitative understanding of the molecular behaviors in supercritical fluids and the physical mechanisms governing their manifestation in fluid properties. However, the current understanding of intermolecular interactions and their physical manifestation in real-fluid effects remains seriously limited, which is fundamentally hampering the widespread application of supercritical combustion.

Today, two pragmatic methods have been developed to model supercritical combustion in homogeneous mixtures. Without considering transport phenomena, the two methods have been focusing on the real-fluid effects on reaction kinetics and thermodynamic properties.

The first type of method artificially modifies the fuel chemistry model (mostly the reaction kinetic parameters therein) in order to match the model predictions with fundamental experiments, while still assuming ideal gas behaviors, e.g., the non-Arrhenius expression in a syngas chemistry model proposed by Sivaramakrishnan et al. [6] and the propane chemistry model with brutal modifications to the reaction kinetic parameters therein [7]. Apparently, these modifications lack physical justification since the thermodynamic and kinetic treatments of the mixture assumed ideal gas behavior. The modified models will most likely fail at other supercritical conditions beyond the trained conditions.

The second type of method relies on using a real-fluid equation of state (EoS), which is often incorporated into the reactor model instead of the fuel chemistry model, to predict the thermodynamic properties and, less often, chemical kinetics of the real-fluids during supercritical combustion. The famous Van der Waals (VdW) EoS was first proposed in 1873 [8], where two empirical constants are used to represent the level of real-fluid behavior. The VdW EoS was later extended and improved to many cubic EoSs, including the Redlich- Kwong (RK) EoS [9], and its modified forms, the Redlich- Kwong (RK) EoS [10] and the Peng-Robinson (PR) EoS [11]. These EoSs have been successfully implemented in supercritical combustion modeling in the past decades, such as the implementation of the PR EoS in the Chemkin Real Gas package [12, 13] and GRI-Mech 1.2 [14], and, more recently, the implementation of the RK EoS in Cantera [15]. Despite the value and convenience of these cubic EoSs, they are semi-empirical in essence and lack physical significance in representing the intermolecular interactions in real fluids. Moreover, as the parameters in the EoS are mostly derived from limited high-pressure experiments, cubic EoS might become inaccurate or even fail at wider pressure conditions.

Aware of the insufficiencies of empirical real-fluid EoSs, the Virial EoS was later developed based on the real-fluid partition function theory in statistical mechanics [16], where fluid non-ideality is quantified using Virial coefficients in the Virial EoS. Note that a Virial EoS can have different orders. The $N^{th}$-order Virial EoS can be expressed as:

$$Z = \frac{P\bar{v}}{RT} = 1 + \frac{B_2(T)}{\bar{v}} + \cdots + \frac{B_N(T)}{\bar{v}^{N-1}}$$

where, $P$ is the pressure, $T$ is the temperature, $\bar{v}$ is the molar volume, $R$ is the universal gas constant, $Z$ represents the compressibility factor, and $B_2, B_3, \ldots, B_N$ are the second, third, …, and $N^{th}$-order Virial coefficients which represent two-body, three-body, …, and N-body intermolecular interactions, respectively. These Virial coefficients physically represent the intermolecular interactions in the fluid, which are computed from pre-determined intermolecular potentials. Hence, the accuracy of the Virial EoS is dictated by the intermolecular potentials adopted for Virial coefficient computation. With such awareness, considerable efforts have been addressed to determine the intermolecular potential for different systems, among which the *ab initio* potential (obtained via quantum mechanics computation) has been the most accurate [17, 18].

Efforts to determine *ab initio* potentials can be dated back to 1965 [19]. Due to the expensive nature of *ab initio* calculations (computational time increases exponentially with the number of bodies considered), to date multi-body pure-substance and cross-body potentials have only been computed for a limited set of species [20]. On the other hand, the calculation of Virial coefficients from intermolecular potentials is also non-trivial, requiring the solving of cluster integrals [21] that is computationally complicated itself. As a result, *ab initio* Virial coefficients have been derived for even less species, e.g., $H_2O$ [22], $CO_2$ [17], $N_2$ [23], and Ar [24]. The available Virial coefficients to date are certainly inadequate for supercritical combustion modeling, as the simplest combustion system (i.e., $H_2$-$O_2$ combustion) will involve over 10 species (radicals included). Furthermore, determining the Virial EoS for a combustion system requires deviation of Virial coefficients for both the pure species in the system and cross-body Virial coefficients between different species. Such a dilemma has made the application of *ab initio* potentials and high-order Virial EoS in supercritical combustion modeling infeasible. This pessimistic situation has apparently persisted to the present day, despite the possibility that *ab initio* Virial EoS might achieve a much better representation of supercritical fluids with regard to their properties and behaviors. It is also vital to note that the fidelity of Virial EoS can be tremendously undermined if it is developed based on inaccurate and unphysical intermolecular potentials or if the order of Virial EoS is low. In fact, this has been demonstrated in a recent study [25], where a reduced form of Virial EoS ($2^{nd}$-order) derived based on simplified, semi-empirical potentials was used to simulate the real-fluid impacts on $H_2$ oxidation under supercritical conditions. As the three-, multi- and cross-body interactions and non-ideal blending behaviors were completely overlooked, among many other critical limitations, significant errors were introduced into the predicted real-fluid properties.

The role of radicals in contributing to system real-fluid effects might also be of significance as they are abundantly formed in supercritical combustion processes, or any other supercritical reacting systems. For instance, in homogeneous, moderately diluted mixtures with hydrogen-rich fuels, some important and long-lived radicals, such as hydroxyl radical (OH) and hydroperoxyl radical ($HO_2$), can have the highest molar concentration in the mixture during combustion. Despite the fact that supercritical combustion has developed vigorous activity over the last few decades, all existing studies did not consider any intermolecular interactions or real-fluid effects from radicals.

In this article, we present results that (1) confirm the critical role of intermolecular interactions in dictating thermochemical real-fluid behaviors, thus real-fluid effects, by demonstrating the superiority/robustness of *ab initio* high-order Virial EoS in reproducing real-fluid properties, and (2) prove the significance of radicals in contributing to real-fluid effects by demonstrating their impact on system reactivity in supercritical ignition. This discovery is enabled through a newly developed framework coupling *ab initio* intermolecular potentials, high-order mixture Virial EoS, real-fluid blending theories, real-fluid thermodynamics, and real-fluid combustion conservation laws of species, mass and energy. This comprehensive framework provides a unique lens into the nature of intermolecular interactions in supercritical reacting systems where studying different molecular behaviors (e.g., two- to multi-body interactions, radical interactions) using experimental techniques is infeasible. Through a combination of calculations and simulations (with results and methods detailed below), we find that high-order Virial EoS developed based on *ab initio* intermolecular potentials reproduces impressively the properties of real fluids and their mixtures, predicting nearly identical results with the experimental measurements. We further reveal the strong impact of intermolecular interactions on system reactivity from thermochemical perspectives, and conclude with discussions emphasizing the necessity to incorporate these behaviors in modeling supercritical combustion, as well as other dynamic reacting systems involving supercritical fluids.

## Results

To investigate the role of *ab initio* molecular potentials and high-order Virial EoS, we first quantified the compressibility factor ($Z$), which describes the deviation of a real fluid from ideal fluid behavior ($Z=1$ corresponds to an ideal fluid). Four pure substances (namely Ar, $N_2$, $CO_2$ and $H_2O$) were first considered, which are the major diluents or products used in fundamental and practical combustion research. These pure substances, though simple, have different molecular properties such as polarity, atom number, etc., which can provide a comprehensive evaluation of the robustness and accuracy of the new framework. To highlight the superiority of our framework, RK EoS and the $2^{nd}$-order Virial EoS based on empirical molecular potentials were also adopted for comparison, along with the experimental measurements [26-29] over 1-1050 bar (Fig. 1a). We find that compressibility factors computed from *ab initio* high-order Virial EoS (either $3^{rd}$-order or $8^{th}$-order) impressively overlapped with experimental data over all the pressure conditions, strictly reproducing the positive pressure dependence of $N_2$ and Ar, the negative pressure dependence of $H_2O$, and the non-monotonic pressure dependence of $CO_2$. In contrast, the RK EoS provided good agreement with experiments only at pressures below 150 bar, with rapidly deteriorating performance at higher pressures (i.e., underpredicting the compressibility factor). The $2^{nd}$-order Virial EoS based on empirical molecular potentials performed the worst and even failed to capture the qualitative trend in compressibility factors for $CO_2$ and $H_2O$, highlighting the critical role of *ab initio* potential in developing Virial EoS. The same results were confirmed with different binary mixtures (Fig. 1b).

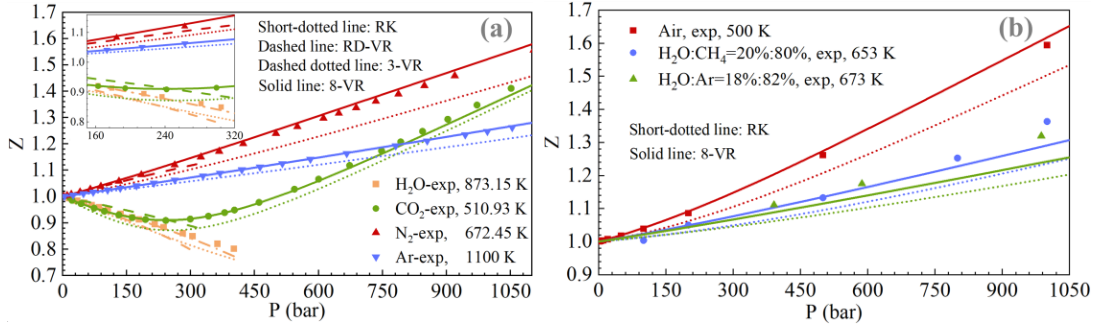

**Figure 1. Compressibility factors calculated using different methods. a, b,** Calculated results for pure Ar, $N_2$, $CO_2$ and $H_2O$ using *ab initio* Virial EoS (the 8th-order: 8-VR, the 3rd-the order: 3-VR), RK EoS, along with experiment data [26] and 2nd-order Virial EoS based on semi-empirical potentials (RD-VR) [16] (**a**); calculated results for binary mixtures using 8-VR and RK EoS, along with experimental date from [27-29] (**b**). *Ab initio* high-order Virial EoS is found to excellently reproduce experimental compressibility factors for all systems, considerably better than cubic EoS and Virial EoS developed with semi-empirical potentials.

We further investigated whether the thermodynamic properties of supercritical fluids can be adequately described with *ab initio* intermolecular potentials. To answer this question, we calculated the molar enthalpy and molar heat capacity of supercritical Ar, $N_2$, $CO_2$ and air (Fig. 2) using real-fluid Departure Function theories [16], where a set of thermodynamic Departure Functions for mixtures based on *ab initio* high-order Virial EoS were developed for the first time. Again, to establish quantitative comparisons, calculations were also conducted using Departure Functions based on RK EoS. The analysis revealed that real-fluid molar enthalpies and molar heat capacities computed from the *ab initio* 6th-order Virial EoS agreed excellently with experiments, whereas the results computed based on RK EoS constantly underpredicted the experiments. RK EoS introduced significant errors at pressures higher than 100 and 150 bar for enthalpy and specific heat, respectively. In almost all cases, the thermodynamic properties computed from the *ab initio* 6th-order Virial EoS are indistinguishable from the experiments, revealing that *ab initio* intermolecular interactions accurately mediate the real-fluid effects on thermodynamic properties.

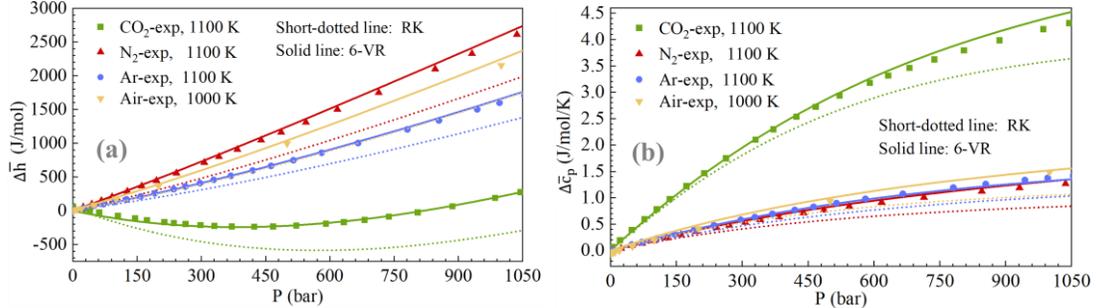

**Figure 2. Real-fluid departure of thermodynamic properties calculated using different methods. a, b,** Calculated results for the molar enthalpy of Ar, $N_2$, $CO_2$ and air (**a**); Calculated results for the molar heat capacity at constant pressure of Ar, $N_2$, $CO_2$ and air (**b**). Results are calculated by Departure Functions [16] newly derived based on *ab initio* 6th-order Virial EoS and RK EoS, along with the experiments from [26, 27]. The comparisons show that *ab initio* intermolecular interactions accurately mediate the real-fluid effects on thermodynamic properties.

The next question we investigated was whether *ab initio* high-order Virial EoS can adequately capture the non-ideal blending behavior between real fluids, an essential feature for supercritical combustion mixtures where hundreds and thousands of real-fluid species blend with each other in a dynamic manner. This was achieved by analyzing the species partial molar volume and non-ideal additivity rules in supercritical mixtures. The mixture molar volume, $\bar{v}$, can be directly calculated based on mixture *ab initio* Virial EoS. Alternatively, $\bar{v}$ can also be evaluated based on non-ideal additivity rules, $\bar{\varphi} = \sum_i X_i \bar{\varphi}_i$, and partial molar volume of all the species in the mixture, $\bar{\varphi}_i = \left(\frac{\partial \varphi}{\partial n_i}\right)_{T,P,n_{j(j \neq i)}}$, where $\varphi$ is a mixture extensive property, and $n_i$ and $X_i$ denote the amount and mole fraction, respectively, of component $i$. If the mixture molar volumes computed from the two different methods equal, then the partial molar volumes and the additivity rules of partial properties are proved. This was confirmed in Fig. 3a over all temperatures and pressures. Following this, the partial molar volume of $H_2O$ in supercritical $CO_2$-$H_2O$ mixtures was computed using different methods (Fig. 3b). It is also clear that in all cases and at all conditions, the results computed based on *ab initio* high-order Virial EoS outperformed the other methods, achieving the best agreements with the experiments. These results indicate that *ab initio* intermolecular interactions can physically and accurately represent the real-fluid blending behaviors in supercritical mixtures.

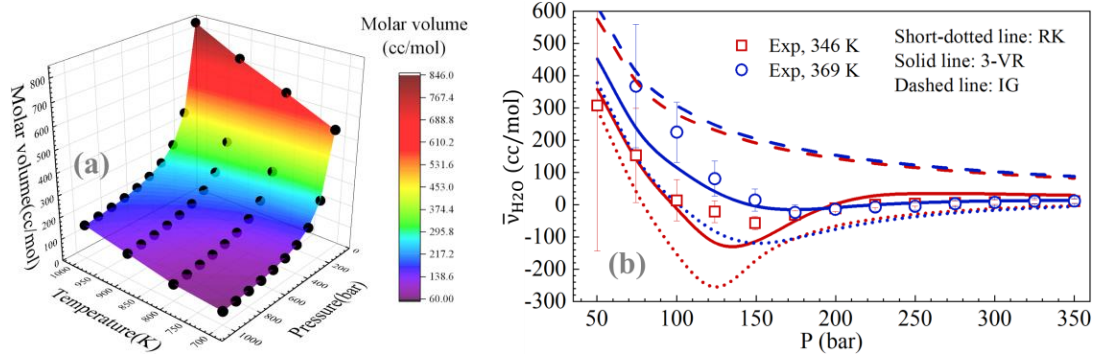

**Figure 3. Real-fluid partial molar properties computed from *ab initio* high-order Virial EoS. a, b,** Comparison between mixture molar volume (surface, computed from mixture *ab initio* 3$^{rd}$-order Virial EoS) and composition-weighted sum of the calculated partial molar volume (solid black circles, computed from species partial molar volume and non-ideal additivity rules, e.g., $\left(\frac{\partial \varphi}{\partial n_i}\right)_{T,P,n_{j(j\neq i)}}$ and $\bar{\varphi} = \sum_i X_i \bar{\varphi}_i$) for a 50:50 mol% $CO_2$-$H_2O$ mixture (**a**); Partial molar volume of $H_2O$ in a 99.7162:0.2838 mol% $CO_2$-$H_2O$ mixture calculated based on the ideal gas (IG) EoS, RK EoS, and the *ab initio* 3$^{rd}$-order Virial (3-VR) EoS, along with experiment data from [31] (**b**). The data in (**a**) prove the derivation of partial molar properties and non-ideal additivity rules based on *ab initio* high-order Virial EoS, and the results in (**b**) show the *ab initio* high-order Virial EoS adequately elucidates species partial molar volume in real mixtures.

The final question we explored was whether or not these real-fluid effects on system properties affect the global reactivity, and if so, how significant the impact is and what are the contributions from different intermolecular interactions. To answer these questions, we simulated the autoignition process for a syngas/$O_2$/$CO_2$ mixture based on ideal gas law and the *ab initio* high-order Virial EoS under three settings: (i) different levels of intermolecular interactions (2-body, 3-body, 4-body and 8-body) were considered; (ii) with and without considering the real-fluid effects of radicals; and (iii) over wider pressure and temperature conditions. The results in Fig. 4a reveal that the real-fluid effects greatly promote the mixture autoignition reactivity, advancing the autoignition delay time ($\tau$), a commonly adopted index for mixture reactivity, by more than 35% at the studied condition. The real-fluid effects are predominantly contributed by 2-body intermolecular interactions, less by 3-body intermolecular interactions (advancing ignition further by ~5 μs, accounting for ~15% of total real-fluid effect), and negligibly by 4-body and above interactions, indicating that 3-body intermolecular interactions are already sufficient to represent thermochemically the real-fluid effects in supercritical combustion modeling. Interestingly, the radicals pose a reversed impact on autoignition reactivity (Fig. 4b), retarding the autoignition delay time by ~10 μs. This magnitude is significant, considering the total real-fluid effect is at ~35 μs. At wide temperatures and pressures (Fig. 4c), real-fluid effects on autoignition reactivity vary with a strong linear dependence on pressure and a somewhat non-linear dependence on temperature, which can lead to a 59.2% increase in autoignition reactivity (i.e., at 1000 bar and 1800 K). Overall, these results indicate that intermolecular interactions considerably influence ignition reactivity in thermochemical manners with radicals playing a significant role.

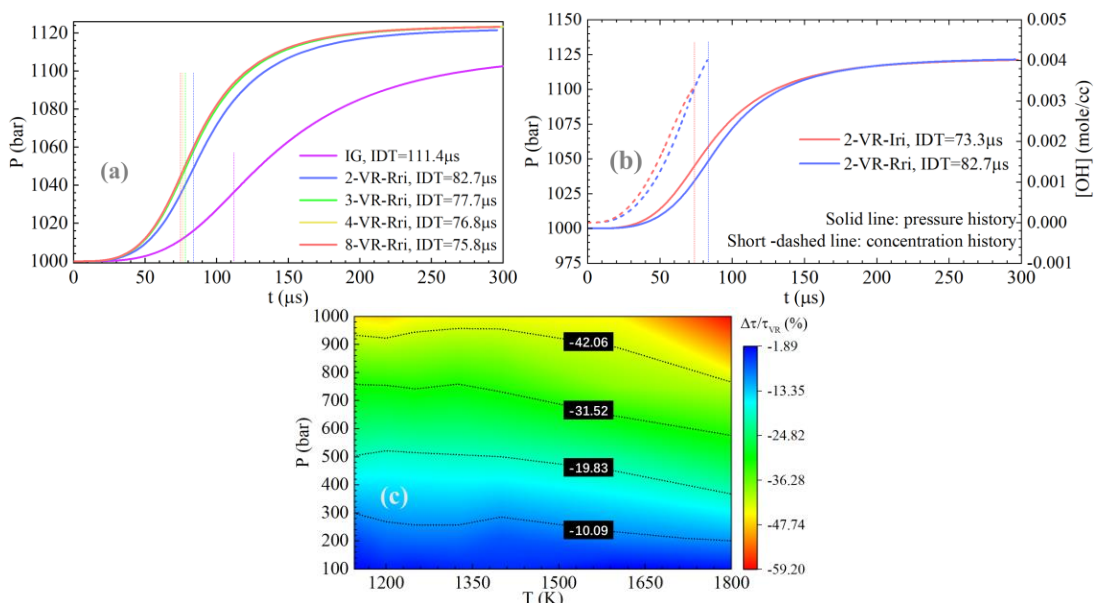

**Figure 4. Contribution of real-fluid effects computed from *ab initio* high-order Virial EoS to autoignition reactivity for a 2.85:0.15:1.5:95.5 mol% $H_2$-CO-$O_2$-$CO_2$ mixture. a-c,** Simulated pressure-time histories at 1000 bar and 1200 K using different orders of *ab initio* Virial EoS, along with the results from the ideal gas simulation (IG) (**a**); and using *ab initio* 2$^{nd}$-order Virial EoS with (2-VR-Rri) and without (2-VR-Iri) considering radical real-fluid effects along with the OH molar concentration (**b**); real-fluid effects on autoignition reactivity calculated by the *ab initio* 8$^{th}$-order Virial EoS at T=1145-1800 K and P=100-1000 bar, quantified as $\Delta\tau/\tau_{VR} = (\tau_{IG} - \tau_{VR})/\tau_{VR}$ where $\tau_{IG}$ and $\tau_{VR}$ are the ignition delay time computed using ideal gas simulation and supercritical simulation via the *ab initio* 8$^{th}$-order Virial EoS, respectively (**c**). Vertical dotted lines in (a) and (b) mark the onset of ignition. The data show the major promoting effect of real-fluid effects on autoignition reactivity, and the significant and inhibiting contribution from radicals, which weigh about 30% of total real-fluid effects.

## Discussions

The change in autoignition reactivity due to intermolecular interactions can be rationalized on the basis of real-fluid thermochemistry and overall chain branching. Thermodynamically, the real-fluid behavior of the system will be dominated by the intermolecular interactions from the most abundant species (e.g., the bath gas $CO_2$ used in Fig. 4) in the mixture that does not chemically participate in reactions. With the system undergoing the same amount of energy change, a real fluid will respond more rapidly in system extensive properties than an ideal fluid. This is confirmed in Figs. 2 and 4 where the pressure (hence temperature) is raised at a higher rate than the ideal mixture. Increased pressure and temperature will certainly promote system reactivity, leading to reduced autoignition delay times. Kinetically, the bath gas can also participate in termolecular reactions either chemically [32] or as a collisional third body [33] to provide activation energy. Such termolecular reactions have been very commonly adopted in modern hydrogen combustion mechanisms [34], e.g., H+H+M=$H_2$+M, O+O+M=$O_2$+M and H+OH+M=$H_2O$+M, where M can be $CO_2$, $O_2$, etc. In general, these pathways will produce at most the same number of radicals as they consume, thus reducing the overall radical branching ratio that leads to a lower system reactivity [35]. However, when multi-body intermolecular interactions between the bath gas species are implemented, strong attractive forces between bath gas molecules are imposed and the possibility of their collision with other radicals decreases. As a result, the suppression effect on system reactivity by termolecular reactions diminishes, leading to an overall increased system ignition reactivity.

The inhibiting effects of radical interactions on system reactivity can also be unraveled from similar perspectives. During combustion, radicals have completely different roles from bath gases, as radicals participate chemically in reactions in an extremely active manner. The rate of reaction is proportional to both the reaction rate coefficient and the molar concentration of the reactants. In supercritical mixtures, the molar concentration of a species can be described as $[X_i]=\frac{X_i P}{ZRT}$, where $Z$ is the compressibility factor (c.f. Fig. 1). As $Z$ is typically above 1 at supercritical conditions, the molar concentration of radicals in a real mixture becomes smaller than that in an ideal mixture, resulting in lower reaction rates. Furthermore, similar to termolecular reactions, reactions with only radicals as reactants can produce at most the same number of radicals as they consume. In fact, most radical-radical reactions in hydrogen combustion are chain termination reactions (e.g., OH+OH=$H_2O$+O and H+OH=$H_2$+O) that reduce global reactivity. With strong radical interactions considered, the radicals are more likely to react with each other rather than participating in other chain propagation or branching reactions. Indeed, the simulations with radical interactions yielded higher total radical mole fraction than the case without radical interactions (c.f. Fig. 4b). The reductions in both radical molar concentration and radical branching ratio eventually lead to the inhibiting impact of radical real-fluid effects on global ignition reactivity.

It is interesting to see a competing mechanism between the real-fluid effects of radicals and bath gases. This competing mechanism governs the overall real-fluid effects from intermolecular interactions, which are most likely to vary greatly with different mixture compositions. For instance, if the $CO_2$ (triatomic, non-polar) in Fig. 4 is replaced with Ar (monatomic, non-polar), $N_2$ (diatomic, non-polar), $H_2O$ (triatomic polar) or their random mixtures, the intermolecular interactions will behave dramatically differently. Accordingly, the real-fluid effects will shift both qualitatively and quantitatively. In fact, preliminary simulations have confirmed the shifts in real-fluid effects with different mixture molecular properties, highlighting the necessity of adequately representing mixture-specific intermolecular interactions in supercritical combustion modeling rather than using empirical or semi-empirical potential functions that lack molecular identity. On the other hand, the relevance of fuel in the mixture cannot be belittled either, as radicals are initiated from fuel molecules. For instance, with the same mole fraction of hydrogen and iso-octane ($C_8H_{18}$) in the mixture, iso-octane will produce up to 9 times higher molar concentration for H-containing radicals than hydrogen in a constant volume environment. As a result, the inhibiting effects from radicals will be amplified, and that the overall real-fluid effects in an iso-octane mixture may be completely different from a hydrogen mixture. Fortunately, the framework proposed herein offers a sufficient means where intermolecular interactions for both bath gases and radicals can be characterized with specific and robust *ab initio* potentials.

Furthermore, the importance of physically describing real-fluid effects from *ab initio* intermolecular interactions need not be restricted to combustion situations, but can be championed more generally. For instance, at altitudes close to the surface, Venus has a mean temperature of 737 K and a pressure of 92 atm [44]. These are above the critical points of the major atmospheric components of Venus (i.e., 96.5% $CO_2$ and 3.5% $N_2$). Under sharp energy deposition events such as solar activities, lightning or meteor strikes, the $CO_2$ can decompose into CO and O, and the produced O can react with $N_2$ to form NO [30, 36]. Following the discussions above, these reacting processes can be significantly enhanced by the intermolecular interactions between $CO_2$ and $N_2$ molecules while suppressed by the radical interactions. These new insights might shed some new light on the historical and prospective atmospheric dynamics on Venus, and help design more reliable Venus probes. Another relevant application is the use of hydrothermal and hydrolytic reactions or oxidative reactions in supercritical water for advanced waste treatment and biomass processing [37-39]. With the presence of supercritical water and supercritical radicals produced from the reactions, the intermolecular $H_2O$ and radical interactions will play an important role in the treatment process. Last but not least, cryogenic supercritical hydrogen storage [40] is a recently arisen application that offers unique advantages over other hydrogen storage techniques. Under leaking events, the cryogenic supercritical hydrogen has been found to react more vigorously with air at cryogenic temperatures than at room temperatures [41]. This could be a combined promoting effect from cryogenic temperature quantum tunneling [42] and real-fluid intermolecular interactions as discussed above. Leveraging the framework developed in this Article, one can provide critical information to inform the safety codes and standards of hydrogen infrastructure with regard to cryogenic supercritical hydrogen storage.

Finally, from the viewpoint of chemical kinetic mechanism development, there is no doubt that real-fluid effects should be fully taken into account when validating chemical kinetic mechanisms against fundamental combustion experiments, particularly at high-pressure conditions where existing mechanism validation studies have been falling short. Taking Fig. 4 as a reference, excluding the thermochemical real-fluid effects can lead to a 60% error in autoignition simulations. Considering the typical measurement uncertainty of ignition delay times, which is on the order of 15% [43], a 60% error in simulation results could disavow any meaningful validation work that did not consider real-fluid effects. In addition to the real-fluid effects that are characterized thermochemically (i.e., via thermodynamics and chemical equilibria) in this Article, detailed consideration of the real-fluid effects on chemical kinetics (e.g., via statistical rate theory) will be required to fully uncover the possible role of real-fluid effects in a homogeneous supercritical reacting environment.

## Methods

We developed a novel framework to determine the high-order mixture Virial EoS based on ab initio intermolecular potentials. The logic is illustrated in Fig. 5, with the key steps discussed as follows:
(1) Calculating pure-substance high-order Virial coefficients. Except for those having been calculated from the using *ab initio* methods, for minor substances including radicals, the $2^{nd}$- and $3^{rd}$-order Virial coefficients were computed from Tsonopoulos [44] and Orbey-Vera [45] correlations respectively, which are based on the theorem of corresponding states. Through these correlations, Virial coefficients were calculated based on temperature, critical point properties, eccentric factors, and the dipole moment. If the critical points are unknown (e.g., for radicals), the Joback group contribution method [46], which regards thermal properties as a result of added contributions of individual functional groups, was employed to estimate critical temperatures and pressures.
(2) Calculating the cross-body Virial coefficients. The 2nd- and 3rd-order cross-body Virial coefficients were determined using the Tsonopoulos correlations [44]. Whenever necessary, critical point properties, eccentric factors, and the dipole moment of a binary mixture were predicted by non-ideal mixing rules [47]. The 3rd-order cross-body Virial coefficients were computed using Chueh-Prausnitz [48] correlations.
(3) Calculating high-order (>3) cross-body Virial coefficients. To date, there are no theories for calculating >3rd-order cross-body Virial coefficients, which is the reason why the high-order Virial EoS has barely been used to predict mixture properties. Inspired by general mean value calculation methods including arithmetic, linear, and harmonic rules [48], we proposed a new combining method based on the arithmetic mean rule. For instance, the expression of the $(r+s)^{th}$-order cross-body Virial coefficient for a binary $R$-$S$ system is:

$$B_{rs} = \frac{rB_{r+s,0} + sB_{0,r+s}}{r+s} \qquad r,s \geq 0 \qquad (1)$$

where, $B_{r+s,0}$ and $B_{0,r+s}$ are the $(r+s)^{th}$-order pure-substance Virial coefficients of substances R and S respectively. This new combining method was extended to any multi-component system.
(4) Calculating high-order mixture Virial coefficients via non-ideal mixing rules [47], the computed pure-substance high-order Virial coefficients, and computed cross-body high-order Virial coefficients. Based on the mixture Virial coefficient, the corresponding *ab initio* high-order mixture Virial EoS and compressibility factor (i.e., results in Fig. 1) were obtained.

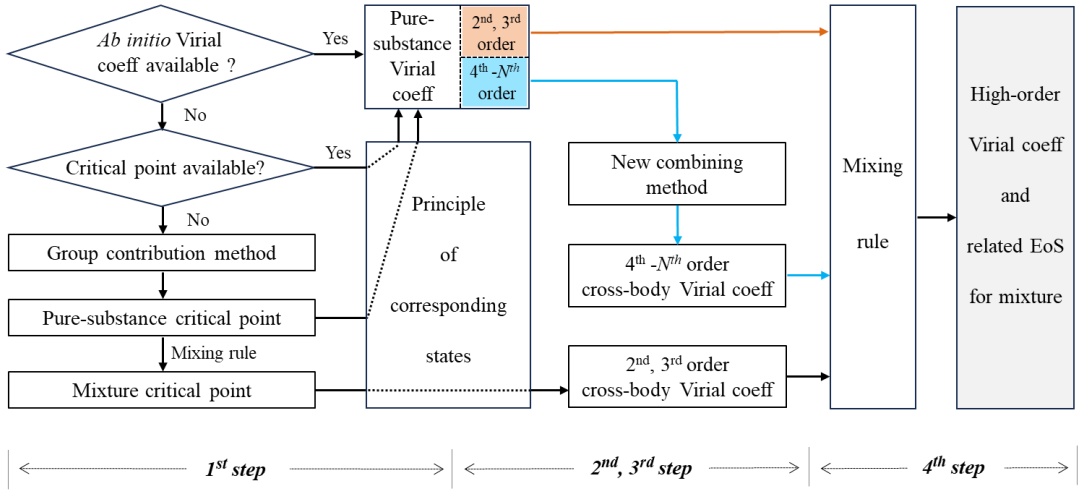

**Figure 5. The logic of determining high-order mixture Virial EoS based on *ab initio* intermolecular interactions.**

Real-fluid thermodynamic properties (i.e., results in Fig. 2) were calculated as the sum of the ideal ones (calculated using NASA polynomials in CANTERA [49]) and respective real-fluid departure, which were determined from Departure Function theories that were newly developed based on high-order mixture Virial EoS. The related departure functions based on the $N^{th}$-order Virial EoS for enthalpy and heat capacity are given as follows:

$$\Delta \bar{h} = RT \sum_{k=2}^{N} \frac{1}{\bar{v}^{(k-1)}} \left( B_k - \frac{1}{k-1} T B'_{k,T1} \right) \qquad (2)$$

$$\Delta \bar{c}_p = -R - R \sum_{k=2}^{N} \frac{2TB'_{k,T1} + T^2 \bar{v} B''_{k,T2}}{(k-1)\bar{v}^{(k-1)}} + \frac{1}{1 + \sum_{k=2}^{N} \frac{kB_k}{\bar{v}^{(k-1)}}} R \left( 1 + \sum_{k=2}^{N} \frac{B_k + TB'_{k,T1}}{\bar{v}^{(k-1)}} \right)^2 \qquad (3)$$

$$B'_{k,T1} = \frac{\partial B_k}{\partial T} \qquad B''_{k,T2} = \frac{\partial^2 B_k}{\partial T^2} \qquad (4)$$

where $\bar{h}$ is the molar enthalpy, $\bar{c}_p$ is the molar heat capacity at constant pressure, and Δ represents a departure.

For component $i$ in a mixture, the final expressions of partial molar volume $\bar{v}_i$ and partial molar enthalpy $\bar{h}_i$ (i.e., results in Fig. 3) were also newly derived based on the $N^{th}$-order Virial EoS:

$$\bar{v}_i = \frac{1}{\frac{1}{\bar{v}} + \sum_{k=2}^{N} \frac{kB_k}{\bar{v}^k}} \left(1 + \sum_{k=2}^{N} \frac{B'_{k,Xi}}{\bar{v}^{(k-1)}}\right) \tag{5}$$

$$\bar{h}_i = \Delta\bar{h} + \bar{h}_{i,p}^{IG} + P(\bar{v}_i - \bar{v}) - RT^2 f \tag{6}$$

$$f = \sum_{k=2}^{N} \frac{\bar{v}B''_{k,Xi,T1} - [(k-1)\bar{v}_i + \bar{v}]B'_{k,T1}}{(k-1)\bar{v}^k} \tag{7}$$

$$B'_{k,Xi} = \frac{\partial B_k}{\partial X_i} \qquad B''_{k,Xi,T1} = \frac{\partial B_k}{\partial X_i \partial T} \tag{8}$$

where, $\bar{h}_{i,p}^{IG}$ is the ideal molar enthalpy of the pure substance $i$.

The *ab initio* high-order Virial EoS, real-fluid thermodynamics and real-fluid partial properties were incorporated to solve the conservation laws of species, mass and energy for homogeneous real-fluid combustion in an open control volume with low Mach numbers, following the logic depicted in Fig. 6. The simulations (i.e., results in Fig. 4) were conducted using CANTERA [49] and the GRI-Mech 3.0 [50].

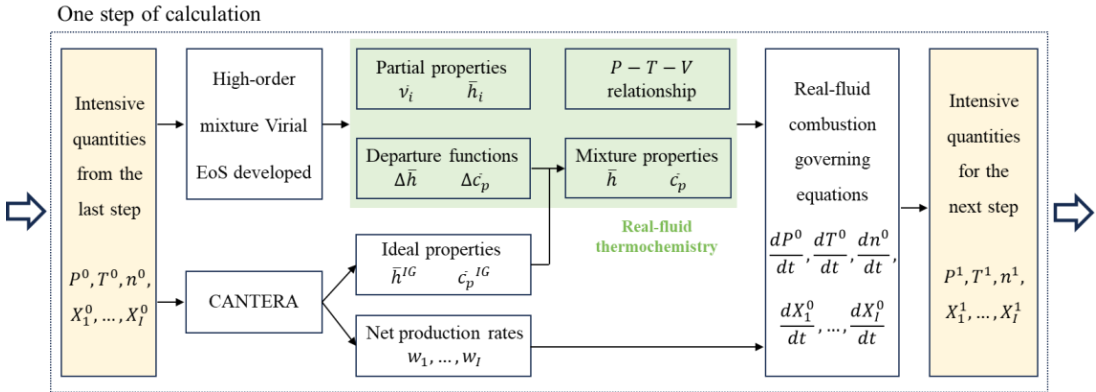

**Figure 6. Supercritical combustion modeling framework based on the *ab initio* high-order Virial EoS.**

Details of the methodology, including deviations of the theories and governing equations, are provided in the Supplementary Material.

**Data availability.** The data shown in Figs. 1-4 are available upon request from the corresponding author. These results can be reproduced using the governing equations derived in the Supplementary Material and publicly available codes and mechanisms, such as CANTERA and GRI-Mech 3.0.

## References

[1] Burke, M. P., Chaos, M., Ju, Y., Dryer, F. L., & Klippenstein, S. J. (2012). Comprehensive H2/O2 kinetic model for high-pressure combustion. International Journal of Chemical Kinetics, 44(7), 444-474.
[2] Li, M., Wu, H., Zhang, T., Shen, B., Zhang, Q., & Li, Z. (2020). A comprehensive review of pilot ignited high pressure direct injection natural gas engines: Factors affecting combustion, emissions and performance. Renewable and Sustainable Energy Reviews, 119, 109653.
[3] Bencivenga, F., Cunsolo, A., Krisch, M., Monaco, G., Orsingher, L., Ruocco, G., ... & Vispa, A. (2007). Structural and collisional relaxations in liquids and supercritical fluids. Physical review letters, 98(8), 085501.
[4] Liong, K. K., Wells, P. A., & Foster, N. R. (1991). Diffusion in supercritical fluids. The Journal of Supercritical Fluids, 4(2), 91-108.
[5] Carlès, P. (2010). A brief review of the thermophysical properties of supercritical fluids. The Journal of Supercritical Fluids, 53(1-3), 2-11.
[6] Sivaramakrishnan, R., Comandini, A., Tranter, R. S., Brezinsky, K., Davis, S. G., & Wang, H. (2007). Combustion of CO/H2 mixtures at elevated pressures. Proceedings of the Combustion Institute, 31(1), 429-437.
[7] Xia, W., Peng, C., Zou, C., Liu, Y., Lu, L., Luo, J., ... & Shi, H. (2020). Shock tube and modeling study of ignition delay times of propane under O2/CO2/Ar atmosphere. Combustion and Flame, 220, 34-48.
[8] Van Der Waals, J. D., & Rowlinson, J. S. (2004). On the continuity of the gaseous and liquid states. Courier Corporation.
[9] Redlich, O., & Kwong, J. N. (1949). On the thermodynamics of solutions. V. An equation of state. Fugacities of gaseous solutions. Chemical reviews, 44(1), 233-244.

[10] Soave, G. (1972). Equilibrium constants from a modified Redlich-Kwong equation of state. Chemical engineering science, 27(6), 1197-1203.

[11] Peng, D. Y., & Robinson, D. B. (1976). A new two-constant equation of state. Industrial & Engineering Chemistry Fundamentals, 15(1), 59-64.

[12] Schmitt, R. G., Butler, P. B., & French, N. B. (1994). Chemkin real gas: a Fortran package for analysis of thermodynamic properties and chemical kinetics in nonideal systems. University of Iowa.

[13] Tang, W., & Brezinsky, K. (2006). Chemical kinetic simulations behind reflected shock waves. International journal of chemical kinetics, 38(2), 75-97.

[14] Petersen, E. L., Davidson, D. F., & Hanson, R. K. (1999). Kinetics modeling of shock-induced ignition in low-dilution $CH_4/O_2$ mixtures at high pressures and intermediate temperatures. Combustion and flame, 117(1-2), 272-290.

[15] Kogekar, G., Karakaya, C., Liskovich, G. J., Oehlschlaeger, M. A., DeCaluwe, S. C., & Kee, R. J. (2018). Impact of non-ideal behavior on ignition delay and chemical kinetics in high-pressure shock tube reactors. Combustion and Flame, 189, 1-11.

[16] Hirschfelder, J. O., Curtiss, C. F., & Bird, R. B. (1964). The molecular theory of gases and liquids. John Wiley & Sons.

[17] Hellmann, R. (2017). Nonadditive three-body potential and third to eighth virial coefficients of carbon dioxide. The Journal of Chemical Physics, 146(5).

[18] Hellmann, R. (2014). Ab initio potential energy surface for the carbon dioxide molecule pair and thermophysical properties of dilute carbon dioxide gas. Chemical Physics Letters, 613, 133-138.

[19] Kohn, W., & Sham, L. J. (1965). Self-consistent equations including exchange and correlation effects. Physical review, 140(4A), A1133.

[20] Guo, H., Worth, G., & Domcke, W. (2021). Quantum dynamics with ab initio potentials. The Journal of Chemical Physics, 155(8).

[21] Barker, J. A. (1957). Cluster integrals and statistical mechanics of solutions. Proceedings of the Royal Society of London. Series A. Mathematical and Physical Sciences, 241(1227), 547-553.

[22] Babin, V., Medders, G. R., & Paesani, F. (2014). Development of a "first principles" water potential with flexible monomers. II: Trimer potential energy surface, third virial coefficient, and small clusters. Journal of chemical theory and computation, 10(4), 1599-1607.

[23] Gokul, N., Schultz, A. J., & Kofke, D. A. (2021). Properties of supercritical $N_2$, $O_2$, $CO_2$, and $NH_3$ mixtures from the virial equation of state. AIChE Journal, 67(3), e17072.

[24] Jäger, B., Hellmann, R., Bich, E., & Vogel, E. (2011). Ab initio virial equation of state for argon using a new nonadditive three-body potential. The Journal of chemical physics, 135(8).

[25] Bai, J., Zhang, P., Zhou, C. W., & Zhao, H. (2022). Theoretical studies of real-fluid oxidation of hydrogen under supercritical conditions by using the virial equation of state. Combustion and Flame, 243, 111945.

[26] Bell, I. H., Wronski, J., Quoilin, S., & Lemort, V. (2014). Pure and pseudo-pure fluid thermophysical property evaluation and the open-source thermophysical property library CoolProp. Industrial & engineering chemistry research, 53(6), 2498-2508.

[27] Lemmon, E. W., Jacobsen, R. T., Penoncello, S. G., & Friend, D. G. (2000). Thermodynamic properties of air and mixtures of nitrogen, argon, and oxygen from 60 to 2000 K at pressures to 2000 MPa. Journal of physical and chemical reference data, 29(3), 331-385.

[28] Shmonov, V. M., Sadus, R. J., & Franck, E. U. (1993). High-pressure phase equilibria and supercritical pVT data of the binary water+ methane mixture to 723 K and 200 MPa. The Journal of Physical Chemistry, 97(35), 9054-9059.

[29] Japas, M. L., & Franck, E. U. (1985). High pressure phase equilibria and PVT‐data of the water‐oxygen system including water‐air to 673 K and 250 MPa. Berichte der Bunsengesellschaft für physikalische Chemie, 89(12), 1268-1275.

[30] Ferguson, E. E., & Libby, W. F. (1971). Mechanism for the fixation of nitrogen by lightning. Nature, 229(5279), 37-37.

[31] Deering, C. E., Cairns, E. C., McIsaac, J. D., Read, A. S., & Marriott, R. A. (2016). The partial molar volumes for water dissolved in high-pressure carbon dioxide from T=(318.28 to 369.40) K and pressures to p= 35 MPa. The Journal of Chemical Thermodynamics, 93, 337-346.

[32] Burke, M. P., & Klippenstein, S. J. (2017). Ephemeral collision complexes mediate chemically termolecular transformations that affect system chemistry. Nature chemistry, 9(11), 1078-1082.

[33] Lindemann, F. A., Arrhenius, S., Langmuir, I., Dhar, N. R., Perrin, J., & Lewis, W. M. (1922). Discussion on "the radiation theory of chemical action". Transactions of the Faraday Society, 17, 598-606.

[34] Ó Conaire, M., Curran, H. J., Simmie, J. M., Pitz, W. J., & Westbrook, C. K. (2004). A comprehensive modeling study of hydrogen oxidation. International journal of chemical kinetics, 36(11), 603-622.

[35] Burke, M. P., & Klippenstein, S. J. (2017). Ephemeral collision complexes mediate chemically termolecular transformations that affect system chemistry. Nature chemistry, 9(11), 1078-1082.

[36] Cowing, K. (2023). Isotopic Constraints On Lightning As A Source Of Fixed Nitrogen In Earth's Early Biosphere. https://astrobiology.com/2023/05/isotopic-constraints-on-lightning-as-a-source-of-fixed-nitrogen-in-earths-early-biosphere.html

[37] Brunner, G. (2009). Near critical and supercritical water. Part I. Hydrolytic and hydrothermal processes. The Journal of Supercritical Fluids, 47(3), 373-381.

[38] Brunner, G. (2009). Near and supercritical water. Part II: Oxidative processes. The Journal of supercritical fluids, 47(3),

382-390.

[39] Kruse, A. (2009). Hydrothermal biomass gasification. The Journal of Supercritical Fluids, 47(3), 391-399.

[40] Song, Z., Xu, J., & Chen, X. (2024). Design and optimization of a high-density cryogenic supercritical hydrogen storage system based on helium expansion cycle. International Journal of Hydrogen Energy, 49, 1401-1418.

[41] Panda, P. P., & Hecht, E. S. (2017). Ignition and flame characteristics of cryogenic hydrogen releases. International journal of hydrogen energy, 42(1), 775-785.

[42] Gao, L. G., Zheng, J., Fernández-Ramos, A., Truhlar, D. G., & Xu, X. (2018). Kinetics of the methanol reaction with OH at interstellar, atmospheric, and combustion temperatures. Journal of the American Chemical Society, 140(8), 2906-2918.

[43] Büttgen, R. D., Preußker, M., Kang, D., Cheng, S., Goldsborough, S. S., Issayev, G., ... & Heufer, K. A. (2024). Finding a common ground for RCM experiments. Part B: Benchmark study on ethanol ignition. Combustion and Flame, 262, 113338.

[44] Tsonopoulos, Constantine. An empirical correlation of second virial coefficients. AIChE Journal 20.2 (1974): 263-272.

[45] Orbey, H., & Vera, J. H. (1983). Correlation for the third virial coefficient using Tc, Pc and ω as parameters. AIChE Journal, 29(1), 107-113.

[46] Joback, K. G., & Reid, R. C. (1987). Estimation of pure-component properties from group-contributions. Chemical Engineering Communications, 57(1-6), 233-243.

[47] Poling, B. E., Prausnitz, J. M., & O'connell, J. P. (2001). The properties of gases and liquids (Vol. 5). New York: Mcgraw-hill.

[48] Orentlicher, M., & Prausnitz, J. M. (1967). Approximate method for calculating the third virial coefficient of a gaseous mixture. Canadian Journal of Chemistry, 45(4), 373-378.

[49] Goodwin, D. G., Moffat, H. K., & Speth, R. L. (2018). Cantera: An object-oriented software toolkit for chemical kinetics, thermodynamics, and transport processes.

[50] Smith, G. P. (1999). GRI-Mech 3.0. http://www.me.berkley.edu/gri_mech/.

## Acknowledgments

This material is based on work supported by the Research Grants Council of Hong Kong Special Administrative Region, China, under PolyU P0046985 for ECS project funded in 2023/24 Exercise, and by the Natural Science Foundation of Guangdong Province under 2023A1515010976.

## Author contributions

M.W. and S.C. conceived and designed the framework and performed the calculations, with technical inputs from R.T., X.R. and Y.C. All authors contributed to writing the paper.

## Competing interests

The authors declare no competing interests.

## Additional information

The article and Supplementary Information include all source data and materials required to reproduce this work. Correspondence and requests for materials should be addressed to S.C.